\documentclass[11pt,twoside]{article1}
\usepackage{doublespace}
\usepackage{note2}

\usepackage{epsfig,amsmath,latexsym,amssymb}

\usepackage{psfig}

\usepackage{verbatim}

\newcommand{\nc}{\newcommand}
\nc{\rnc}{\renewcommand}

\nc{\smfrac}[2]{\mbox{$\frac{#1}{#2}$}}
\nc{\ox}{\otimes}
\nc{\dg}{\dagger}
\def\ot{\otimes}
\nc{\eq}[1]{Eq.~(\ref{eq:#1})}
\nc{\eqs}[2]{Eqs.~(\ref{eq:#1}) and (\ref{eq:#2})}
\nc{\eqm}[2]{Eqs.~(\ref{eq:#1})--(\ref{eq:#2})}

\rnc{\sec}[1]{Sec$.\,$\ref{sec:#1}}

\def\lbm{ \left[\rule{0pt}{2.1ex}\right. }
\def\rbm{ \left.\rule{0pt}{2.1ex}\right] }

\def\lbL{ \left[\rule{0pt}{2.4ex}\right. \!}
\def\rbL{ \!\left.\rule{0pt}{2.4ex}\right] }

\def\non{\nonumber}
\rnc\ss{\hspace*{0.1ex}}
\nc\ms{\hspace*{-0.1ex}}

\def\Pr{{\rm Pr}}
\def\Tr{{\rm Tr}}
\def\be{\begin{eqnarray}}
\def\ee{\end{eqnarray}}
\def\bea{\\[-0.5ex] \begin{eqnarray}}
\def\eea{\end{eqnarray} \hfill\\[0.7ex]}
\def\r{\varrho}
\def\ra{\rightarrow}

\def\a{\alpha}
\def\e{\epsilon}
\def\k{\kappa}
\def\G{\Gamma}
\def\s{\sigma}
\def\sr{s.r.$\;$}

\def\SD{\mbox{SD}}

\def\A{{\rm A}}
\def\B{{\rm B}}
\def\E{{\rm E}}
\def\I{{\rm I}}

\def\P{{\cal P}}
\def\S{{\cal S}}
\def\CM{{\cal M}}
\def\CE{{\cal E}}

\rnc{\AA}{{\mathbb A}}
\nc{\Q}{{\mathbb Q}}
\rnc{\H}{{\mathbb H}}
\nc{\BB}{{\mathbb B}}
\nc{\EE}{{\mathbb E}}
\nc{\XX}{{\mathbb X}}
\nc{\UU}{{\mathbb U}}

\def\>{\rangle}
\def\<{\langle}
\def\ot{{\otimes}}
\renewcommand\:{\! : \!}
\nc{\mPr}[1]{
	\begin{array}{c}
	\rule{0pt}{0.0ex}\mbox{\rm Pr} \\ \raisebox{0ex}{\scriptsize $#1$}
	\end{array} \!\!
}
\rnc{\mPr}[1]{
	\raisebox{1.5ex}\mbox{\rm Pr} \hspace{-2ex} 
	\raisebox{-1.7ex}{\scriptsize $#1$}~
}

\nc{\mrho}{\raisebox{0.15ex}{$\rho $}}

\begin{document}

\title{The Universal Composable Security of Quantum Key Distribution}


%
%
%
%
%

\author{\small{Michael Ben-Or$^{1,4,6}$, Micha{\l} Horodecki$^{2,6}$,
Debbie$\;$W.$\;$Leung$^{3,4,6}$, Dominic Mayers$^{3,4}$, and Jonathan
Oppenheim$^{1,5,6}$} \\[1ex]
{\footnotesize
benor\@@cs.huji.ac.il, fizmh\@@univ.gda.pl, wcleung\@@cs.caltech.edu, 
dmayers\@@cs.caltech.edu, \& J.Oppenheim\@@damtp.cam.ac.uk 
\\[1ex]
{\it $^1$Institute of Computer Science, The Hebrew University, Jerusalem,
Israel \\
$^2$Institute of Theoretical Physics and Astrophysics, University of
Gda\'nsk, Poland \\
$^3$Institute for Quantum Information, California Institute of
Technology, Pasadena, USA \\
$^4$Mathematical Science Research Institute, 
Berkeley, USA \\
$^5$Department of Applied Mathematics and Theoretical Physics, 
University of Cambridge, Cambridge, UK \\
$^6$Isaac Newton Institute, University of Cambridge, Cambridge, UK}} 
\vspace{1ex}}

\date{\today}
 
\maketitle


{\small \begin{quote}
The existing unconditional security definitions of quantum key
distribution (QKD) do not apply to joint attacks over QKD and the
subsequent use of the resulting key.
%
%
In this paper, we close this potential security gap by using a
universal composability theorem for the quantum setting.
We first derive a composable security definition for QKD.
We then prove that the usual security definition of QKD still implies
the composable security definition.  Thus, a key produced in any QKD
protocol that is unconditionally secure in the usual definition can
indeed be safely used, a property of QKD that is hitherto unproven.
We propose two other useful sufficient conditions for composability.
As a simple application of our result, we show that keys generated by
repeated runs of QKD degrade slowly.
\end{quote}}

\setlength{\parskip}{1ex}
\setstretch{0}
\raggedbottom
\thispagestyle{empty}

\section{Introduction}
\label{sec:motivation}

Quantum cryptography differs strikingly from its classical
counterpart.
On one hand, quantum effects are useful in the construction of many
cryptographic schemes.  On the other hand, dishonest parties can also
employ more powerful quantum strategies when attacking cryptographic
schemes.


\vspace*{1ex} 

{\bf The security of quantum key distribution}

One of the most important quantum cryptographic applications is
quantum key distribution (QKD) \cite{Bennett84a,Ekert91a,Bennett92b}. 
The goal of key distribution (KD) is to allow two {\em remote}
parties, Alice and Bob, to share a {\em secret} bit string.
Classically, KD cannot be unconditionally secure (i.e.~secure against
all possible classical attacks) (see \sec{qkd}).  Furthermore, the
security of existing KD schemes is based on assumptions in computation
complexity or limitations of the memory space of the adversary, Eve.
%
%
%
In contrast, QKD is based on an intrinsic property of quantum
mechanics, ``extracting information about an unknown quantum state
inevitably disturbs it,'' \cite{Bennett94a} which allows eavesdropping
activities to be detected in principle.
%
%
Indeed, QKD can be {\em unconditionally secure}, i.e., against Eve whose
capability is only limited by quantum mechanics
\cite{Mayers96,Mayers98a,Lo99b,Biham00,Shor00,Tamaki02,Gottesman01}.
Furthermore, QKD remains secure even if the quantum states are sent
through a noisy quantum channel, as long as the observed error rates
are below certain threshold values.  

In what sense is QKD secure?  
We will describe the assumptions and security definitions more
formally in \sec{qkd}.  
In QKD, Alice and Bob are assumed to start with a small initial key
$K_{i}$ (for authentication purposes). 
They have access to uncorrelated randomness that is not controlled by
Eve.
They may exchange quantum and classical messages in both directions
via channels that are completely under the control of Eve, and may
perform local quantum operations and measurements. 
Based on their measurement outcomes, Alice and Bob either abort QKD or
generate their respective keys $K_\A,K_\B$.
Correspondingly, we say that the QKD test is failed or passed, and the 
events can be described as $M{=}0$ or $M{>}0$, where $M$ is the length 
of the key generated.
%
%
Eve also obtains quantum and classical data (her ``view'' or
``transcript'') from which she extracts classical data $K_\E$ via a
measurement.
%
%
What happens during a specific run of QKD depends on Eve's strategy as
well as the particular outcomes of the coins and quantum measurements 
of all the parties.
However, the security of QKD can still be captured by requiring that
(1) the conditional mutual information $I(K_\E\,{:}\,K_\A,K_\B \, | M)$ 
is negligible and (2) for all eavesdropping strategies with nonnegligible
$\Pr(M{>}0)$, $K_\A$, $K_\B$ are near-uniform and $\Pr(k_\A \,{\neq}
\, k_\B)$ is negligible.
%
%
%
Throughout the paper, we use capitalized letters $K_{\A}$, $K_{\B}$,
$K_\E$, and $M$ to denote the random variables, and uncapitalized
letters to denote specific outcomes.

\vspace*{1ex} 
{\bf The security problem of using QKD}

Proofs of security of QKD (in the sense described above) address all
attacks on the QKD scheme allowed by quantum mechanics. 
The problem is that QKD is {\em not} the only occasion for attack ---
further attack may occur when Alice and Bob use the keys generated.
In particular, Eve may never have made a measurement during QKD to
obtain any $K_\E$.  
Eve's transcript is a quantum state.  She could have delayed
measurements until after more attack during the application, a 
strategy with power that has no classical counterpart.
In other words, security statements in QKD that revolve around
bounding $I(K_\E\,{:}\,K_\A,K_\B \, |M)$ is {\em not applicable} if
the key is to be used!

The limitations of mutual-information-based security statements were
known as a folklore for some time (for example, see Sec.~4.2 in
\cite{Gottesman01}).
One of the earliest known security problems in QKD is the following
\cite{BennettSmolinHarrow}:
QKD requires a key for authentication, which may in turns come from
a previous round of QKD.  Since each run of QKD is slightly
imperfect, repeated QKD produce less and less secure keys.  A
conclusive analysis on the degradation has been evasive, since joint
attacks over all runs of QKD have to be considered.

As it turns out, there are many other occasions in which joint attacks
on QKD and the subsequent use of the generated key have to be
considered.  For example, suppose Alice and Bob perform QKD to obtain
a key, and then use the key to encrypt quantum states
\cite{Ambainis00,Boykin00}.  Eve eavesdrops during both QKD and
encryption and performs a collective measurement on the two 
eavesdropped states.  It is well-known that such a collective
measurement may yield more {\em accessible information} than the sum
of information obtained in two separate measurements \cite{Peres91}.

Our current study is further motivated by the results in
\cite{DiVincenzo03,HLSW03}, which show that there are ensembles of
quantum states that provide little accessible information on their
own, but can provide {\em much more} information when a little more
{\em classical} data is available.
The extra information can be arbitrarily large compared to both the
initial information and the amount of extra classical data.
Such strange property reveals a new, unexpected, inadequacy of
mutual-information-based statements.  
In particular, in the context of QKD, the usefulness of bounding the
initial accessible information of Eve becomes very questionable, if
Eve delays her measurement until further data is available during
the application of the key --- the security of the key is questionable
even in {\em classical} applications!
%

The goal of the current paper is to study the security of using 
a key generated by QKD, i.e., the composability of QKD. 

\vspace*{1ex} 
{\bf The universal composability approach}

Composability is an active area of research that is concerned with the
security of composing cryptographic primitives in a possibly complex
manner.  
The simplest example is the security of using a cryptographic
primitive as a subroutine in another application.
Our paper will follow the {\em universal composability} approach.  
For a specific task (functionality), a primitive that realizes the
task is said to be universal composable if any application using the
primitive is about as secure as using the ideal functionality.
A security definition that ensures {\em universal} composability was
recently proposed by Canetti \cite{Canetti01}, and was extended to the
quantum setting by some of us \cite{QIP03-1,BM02}.
Such universal composable security definitions are useful because they
are in terms of the ideal functionality only, without reference to the
potential application.  
The security of a complex protocol can then be analyzed in terms of
the security of each individual component in a systematic and 
error-proof manner.  
In the quantum setting, universal composability provides the only
existing systematic technique for analyzing security in the presence
of subtleties including entanglement and collective attacks.
We will see in this paper that universal composability provides the
precise framework for proving the security of using the keys generated
from QKD, a problem that appears intractable at first sight.

We note that an alternative approach to achieve universal
composability in the classical setting was obtained in \cite{BPW04},
with a generalization to the quantum setting studied in
\cite{Unruh04}.

\vspace*{1ex} 
{\bf Main Results}

We have pointed out a serious potential security problem in using the
keys generated from QKD.
We will address the problem in the rest of the paper.
We derive a new security definition for QKD that is universal 
composable.  
The essence is that QKD and certain ideal KD should be
indistinguishable from the point of view of potential adversaries.
Then, we prove that the original mutual-information-based
security definition implies the new composable definition.
Other simple sufficient conditions for the composable security of QKD
will be discussed.
One of these conditions, high singlet-fidelity, has always been an
intermediate step in the widely-used ``entanglement-based'' security
proofs of QKD.  We show that high singlet-fidelity is much more
closely related to composable security than the usual security
definition, and we obtain much better security bounds for known QKD
schemes.
We thus prove the security of using a key generated by QKD in various
ways, and provide simple criteria for future schemes.
As a corollary, we answer the long standing question concerning the
extent of key-degradation in repeated use of QKD
\cite{BennettSmolinHarrow}.

Our work also has non-cryptographic applications in the study of
correlations in quantum systems.  The various security conditions are
tied to correlation measures in quantum systems.  Each derivation for
the composable security for QKD is based on relating a pair of
correlation measures.

\vspace*{1ex} 
{\bf Related work}

Since the current result was initially presented \cite{QIP03-2,QIP04},
various related results were reported.
The composable security of generic classes of QKD schemes were proved
in \cite{Renner04,CRE04}, following a different approach of showing
the composable security of certain privacy amplification procedures
against quantum adversaries \cite{Renner04}.  
These related works share the concerns raised in this paper, with 
results complementary to ours. 

\vspace*{1ex} 
{\bf Organization of the paper} 

We end this section by introducing some basic elements in the quantum
setting.
%
%
We review QKD in \sec{qkd}, stating our definitions and assumptions
more formally.
In \sec{qucompos}, we review the quantum universal composability
theorem.  We will restrict ourselves to the much simpler case
concerning unconditional security.
%
%
We start describing our main results in \sec{qkducdef}, which contains
a derivation of a simple criteria for the universal composable
security for QKD.
In \sec{qkduc}, we prove that the usual security definition for QKD
implies the universal composable security.  
In addition, we demonstrate two other sufficient conditions for
composable security.
One is based on bounding the Holevo information of Eve on the key.
The other is based on bounding the singlet-fidelity in security 
proofs using entanglement-purification.
The latter implies much better security of existing QKD protocols than
is generically implied by the usual security definition.
We conclude with lessons learnt from the current results.  
Frequently used notations and some complicated information theoretic 
quantities are listed in the appendix.  

\vspace*{1ex} {\bf Basic elements of quantum mechanics}


A quantum system or register is associated with a Hilbert space $\H$.
We only consider finite dimensional Hilbert spaces.  Let ${\cal
B}(\H)$ and $\UU(\H)$ denote, respectively, the set of bounded
operators and the unitary group acting on $\H$.  We loosely refer to
the system as $\H$ also.  A composite quantum system is associated
with the tensor product of the Hilbert spaces associated with the
constituent systems.

The state of $\H$ is specified by a positive semidefinite {\em density
matrix} $\rho \in {\cal B}(\H)$ of unit trace.  A density matrix is a
convex combination of rank-$1$ projectors (commonly called {\em pure
states}) and represents a probabilistic mixture of pure states.  Up to
an overall-phase that is not physically observable, pure states can be
represented as vectors in $\H$.  $|\psi\>$ and $|\psi\>\<\psi|$ denote
the vector and rank-$1$ projector respectively.

A measurement $\CM$ on $\H$ is defined by a POVM, which is a
decomposition of the identity into a set of positive semidefinite
operators $\{O_k\}$, i.e., $\sum_k O_k = I$.  If the state is
initially $\rho$, the measurement $\CM$ yields outcome $k$ with
probability $\Tr(O_k \rho)$ and changes the state to $\sqrt{O_k} \rho
\sqrt{O_k}/ \Tr(O_k \rho)$.  $\CM$ is said to be along a basis
$\{|k\>\}$ if $\{O_k\} = \{|k\>\<k|\}$.  Measuring an unknown state
generally disturbs it.

The most general evolution of the state is given by a trace-preserving
completely-positive (TCP) linear map $\CE$ acting on ${\cal B}(\H)$.
Any such $\CE$ can be implemented by preparing a pure state in some
ancillary system $\H'$, applying a joint unitary operator $U \in
\UU(\H \, \ot \, \H')$, and discarding $\H'$ (i.e., a partial trace
over $\H'$).

We mention two distance measures for quantum states.  The first is the
trace distance $\| \mrho_1-\mrho_2 \|_1$ between the density matrices.
It can be interpreted as the maximum probability 
%
of distinguishing between the two states. 
The second measure is the fidelity, $F(\mrho_1, \mrho_2) =
\max_{|\psi_1\>,|\psi_2\>} |\<\psi_1|\psi_2\>|^2$ where $\mrho_{1,2}
\in {\cal B}(\H)$, $|\psi_{1,2}\> \in \H \ot \H'$ are
``purifications'' of $\mrho_{1,2}$ (i.e., $\Tr_{\H'}
|\psi_{1,2}\>\<\psi_{1,2}| = \mrho_{1,2}$), and $\<\cdot|\cdot\>$ is
the inner product in $\H$.
%

We refer our readers to the excellent textbook by Nielsen and Chuang
\cite{Nielsen00bk} for a more comprehensive review of the quantum model
of information processing.

\section{Quantum Key Distribution}
\label{sec:qkd}

The goal of key distribution (KD) is to allow two {\em remote}
parties, Alice and Bob, to share a {\em secret} bitstring such that no
third party, Eve, will have much information about the bitstring.
KD is impossible unless Alice and Bob can identify one another and
detect alterations of their communication.
In other words, the task of {\em message authentication} is necessary 
for KD. 
There are unconditionally secure methods for authenticating a
classical message with a much shorter key \cite{Wegman81}. 
Thus, KD uses authentication as a subroutine, and achieves key 
expansion (producing a key using a much shorter initial key).  

Classically, unconditionally secure KD between two remote parties is
impossible.  Classical physics permits an eavesdropper to have exact
duplicates of all communications in any KD procedure without being
detected.
In contrast, while quantum key distribution (QKD) cannot prevent
eavesdropping, it can detect eavesdropping.  
This allows Alice and Bob to avoid generating compromised keys with
high probability.  The usefulness of QKD is to avoid Alice and Bob
being fooled into having a false sense of security.
%
%
It is worth emphasizing what QKD does not offer.  First, QKD does not
promise to always produce a key, since Eve can cause QKD to be aborted
with high probability with intense eavesdropping. Second, there is a
vanishing but non-zero chance that Eve is undetected, so that one
cannot make simple security statements conditioned on not aborting
QKD.

\vspace*{1ex}
{\bf How and why QKD works, through an example}

Various QKD schemes have been proposed and we only name a few here:
BB84 \cite{Bennett84a}, E91 \cite{Ekert91a}, B92 \cite{Bennett92b},
and the six-state scheme \cite{Bruss98,Bechmann}.
%
%
We illustrate the general features and principles behind QKD by
describing the class of prepare-\&-measure schemes.
Recall that Alice and Bob are given secure local coin tosses. 
%
%
Step 1: Alice first generates a random bitstring, encodes it in some
quantum state $\mrho_\A$, and sends $\mrho_\A$ to Bob through an
insecure quantum channel controlled by Eve.
During this time, Eve can manipulate the message (system $\AA$) in any
way allowed by quantum mechanics.  Eventually, she will have to give
some quantum message $\mrho_\B$ to Bob for QKD to proceed.
Mathematically, Eve's most general operation can be described as
attaching a private system $\EE$ in the state $|0\>\<0|_\E$, applying
a joint unitary operation $U$ to produce a joint state $\mrho = U \,
(\mrho_\A \ot |0\>\<0|_\E) \, U^\dg$, and passing system $\AA$ to Bob
(relabeled as system $\BB$).  Thus, Bob and Eve share the joint state
$\mrho$, and $\mrho_\B := \Tr_\EE \, \mrho$, $\mrho_\E := \Tr_\BB \ss
\mrho$ are their respective reduced density matrices.
Meanwhile, Bob measures $\mrho_\B$ (according to his coin tosses).
Step 2: Bob acknowledges to Alice receipt of the quantum message.
Step 3: Only {\em after} Alice hears from Bob will further classical
discussion be conducted over a public but authenticated channel.  
Step 4: At the end, based on their measurement outcomes and
discussions, Alice and Bob either abort QKD ($m=0$), or generate keys
$K_\A$ and $K_\B$ ($m>0$), and they announce $m$.
Eve will have access to all the classical communication between Alice and
Bob, besides the state $\mrho_\E$.  She can measure $\mrho_\E$ at any
time to obtain a classical string $K_\E$, though it is to her
advantage to wait until after she receive the classical communication.
See Figure~1 for a schematic diagram for the class of
prepare-\&-measure QKD schemes.



\begin{figure}[http]
\setlength{\unitlength}{0.6mm}
\centering
\begin{picture}(140,100)
{\scriptsize
\put(36,75){\framebox{Step 1}}
\put(0,40){\framebox{Step 2}}
\put(0,25){\framebox{Step 3}}
\put(0,10){\framebox{Step 4}}
\put(23,50){\dashbox{0.5}(12,8){Alice}}
\put(22,60){\makebox(10,10){$r_\A$}}
\put(37,60){\makebox(10,10){$\rho_\A$}}
\put(45,65){\line(1,0){10}}
\put(55,65){\vector(1,1){10}}
\put(51,65){\makebox(5,5){$\AA$}}
\put(94,65){\makebox(5,5){$\BB$}}
\put(56,80){\makebox(10,10){$|0\>$}}
\put(65,85){\line(1,0){5}}
\put(65,75){\line(1,0){5}}
\put(70,73){\framebox(10,14){$U$}}
\put(80,75){\line(1,0){5}}
\put(85,85){\makebox(5,5){$\EE$}}
\put(80,85){\line(1,0){15}}
\put(85,75){\vector(1,-1){10}}
\put(95,65){\line(1,0){10}}
\put(104,60){\makebox(10,10){$\rho_\B$}}
\put(95,80){\makebox(10,10){$\rho_\E$}}
\put(103.5,84){\vector(3,-2){6}}
\put(110,75){\framebox(20,8){measure}}
\put(125,88){\dashbox{0.5}(10,8){Eve}}
\put(132,78){\makebox{$k_\E$}}
\put(56,72){\dashbox{0.5}(83,26){}}
\put(120,30){\dashbox{0.5}(11,8){Bob}}
\put(117,60){\makebox(10,10){$r_\B$}}
\put(109,62){\vector(0,-1){5}}
\put(100,50){\framebox(20,8){measure}}
\put(70,40){\makebox(10,5){``received''}}
\put(109,40){\vector(-1,0){68}}
\put(109,39){\vector(-1,0){68}}
\put(109,25){\vector(-1,0){68}}
\put(41,24){\vector(1,0){68}}
\put(70,25){\makebox(10,5){``discussion''}}
\put(109,10){\vector(-1,0){68}}
\put(41,9){\vector(1,0){68}}
\put(70,10){\makebox(10,5){``m''}}
\put(20,5){\dashbox{0.5}(30,65){}}
\put(23,15){\makebox(10,10){$k_\A$}}
\put(23,10){\makebox(10,8){$m$-bit}}
\put(99,5){\dashbox{0.5}(37,65){}}
\put(123,15){\makebox(10,10){$k_\B$}}
\put(123,10){\makebox(10,8){$m$-bit}}
}
\end{picture}
\label{fig:qkd}
\caption{\small Schematic diagram for the class of prepare-\&-measure
QKD schemes.  The classical messages, represented by double lines, are
available to Eve.  Eve can make her measurement any time after step 1.
Dashed boxes represent private laboratory spaces.  Outcomes of Alice
and Bob's local coins are represented by $r_\A, r_\B$.}
\end{figure}
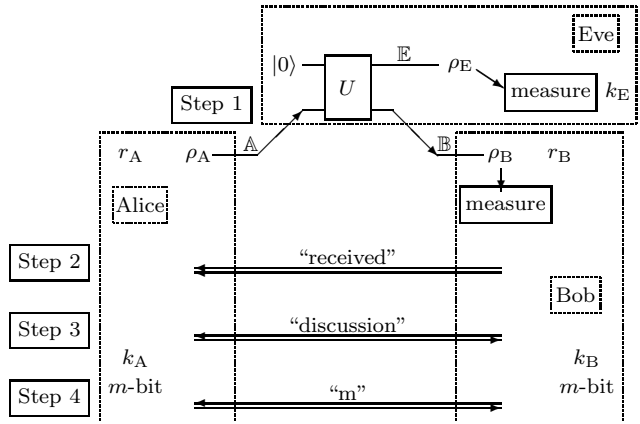

The principle behind QKD is that, in quantum mechanics, one can only
reversibly extract information from an unknown quantum state if the
state is drawn from an orthogonal set \cite{Bennett94a}.
Thus in the prepare-\&-measure scheme described above, if Alice
encodes her message using a random basis chosen from several
nonorthogonal possibilities, and Eve is to obtain any information on
the outcomes of $K_\A$, $K_\B$, then $\mrho_\B \neq \mrho_\A$.
To detect the disparity, Bob measures some of the received qubits (the
``test-qubits'' chosen randomly to avoid Eve tailoring her attack) 
and discusses with Alice to check if his measurement outcomes are
consistent with what Alice has sent.
This intuition can be turned into a provably secure procedure.  Alice
and Bob estimate various error rates on the test-qubits.
If the observed error rates are below certain threshold values, it is
unlikely that the untested qubits have much higher error rates.
Error reconciliation and privacy amplification are applied to extract
bitstrings $k_\A$ and $k_\B$ for Alice and Bob respectively. 
If the observed error rates are above the thresholds, Alice and Bob
abort QKD.
QKD remains secure whether the observed noise is due to natural
channel noise or due to eavesdropping.

\vspace{1ex} 
{\bf General features of any QKD scheme}

There are other QKD schemes besides prepare-\&-measure schemes, for
example, the entanglement-based QKD schemes (see 
\cite{Ekert91a,Lo99b,Deutsch}).
Unless otherwise stated, our discussion applies to all QKD schemes.  
The basic ingredients are still secure local coins, completely
insecure quantum communication, and authenticated public classical
communication between Alice and Bob.  In the most general QKD scheme,
the ingredients may be used in any possible way.  
Alice and Bob still obtain some bitstrings as the output keys, $k_\A$
and $k_\B$, of certain length $m$.
Eve's view is still given by some quantum and classical data, denoted
collectively by $\mrho_{\E,k_\A,k_\B}$, with explicit dependence on
$k_\A$, $k_\B$.  (Her view is a draw from an ensemble.)

We emphasize a limitation in QKD.  It is possible for Eve to be
``lucky,'' for example, to have attacked only the untested qubits, or
to have attacked every qubit without causing inconsistency in Alice
and Bob's measurements.  Thus, it is unlikely, but still possible, for
Eve to have a lot of information on the generated key without being
detected. No QKD protocol can make the promise ``{\em conditioned} on
passing the test, the keys $K_\A$, $K_\B$ will be so-and-so.''
With the above limitation of QKD in mind, there are several approaches 
to a proper security statement. 
The approach that is most commonly used in existing security proofs is
to bound the probability that Alice and Bob generate bitstrings that
are not equal, uniform, or private.
We will use a more compact statement in the following.

Let $n$ be a security parameter in QKD (for example, the number of
qubits transmitted from Alice to Bob).  Fix an arbitrary eavesdropping
strategy.  
The attack induces a distribution $\Pr(M{=}m)$ on the key length $M$.
The average value of $M$ is typically a small fraction of $n$.
The outcome $m$ in a particular run of QKD depends on the outcome of
the coins and measurements by Alice and Bob.
We can assume that $m$ is made {\em public} at the end of QKD.  
Recall $m>0$ if the QKD test is passed and $m=0$ if QKD is aborted.

Let $p_{\rm \ss qkd}^{(m)}$ denote the distribution of $K_\A,K_\B$
generated in QKD conditioned on $|K_\A|=|K_\B|=m$, i.e., 
\bea
	p_{\rm \ss qkd\,}^{(m)}(k_\A,k_\B) = \Pr(K_\A=k_\A,K_\B=k_\B|M=m) \,.
\eea
Let $p_{\rm ideal}^{(m)}$ be the following distribution over two $m$-bit
strings, 
\bea
	\left\{ \begin{array}{llll}
	p_{\rm \ss ideal\,}^{(m)}(l,l) & = & 2^{-m}  
\\[1ex]	p_{\rm \ss ideal\,}^{(m)}(l,l') & = & 0 & {\rm if~} l \neq l' \,.  
	\end{array} \right. 
\eea
Let ${\cal V}$ denote the set of exponentially decaying functions 
of $n$.
%
%
%
With these notations, a simple statement for the security condition
can be made.
\\[2ex]
{\bf Usual security definition for QKD:}\\[0.5ex]
A QKD scheme is said to be secure if the following properties hold for
all eavesdropping strategies. \\[1ex]
{\it $\bullet$~Equality-and-uniformity:} $\exists \mu_1 \in {\cal V}$ 
s.t. 
\bea 
	\sum_{m=0} \Pr(m) \; \big \| \, p_{\rm \ss ideal}^{(m)} -
	p_{\rm \ss qkd}^{(m)} \, \big \|_1 ~\leq~ \mu_1
\label{eq:uniformity}
\eea
{\it $\bullet$~Privacy:} $\exists \mu_2 \in {\cal V}$ s.t.
\bea 
	\sum_{m=0} \Pr(m) \times I(K_\E\,{:}\,K_\A,K_\B \, | \,M=m\,) 
	~\leq~\mu_2
\label{eq:privacyold}
\eea
where $I$ above denotes the mutual information \cite{Cover91a} between
$K_\E$ and $K_\A,K_\B$ conditioned on $M=m$.
Using the equality condition, we only need to focus on $k_\A=:k$ in
\eq{privacyold}.
In particular, 
\\[1ex]
{\it $\bullet$~Privacy:} $\exists \mu_2' \in {\cal V}$ s.t.
\begin{eqnarray}
	\sum_{m=0} \Pr(m) \times I(K_\E\,{:}\,K \, | \,M=m\,) 
	~\leq~\mu_2'
\label{eq:privacy}
\end{eqnarray}
\hfill \\[-0.3ex]
The above security conditions revolve around expressions that can be
interpreted as deviations from the desired properties, averaged over
$m$.
The product in each summand precisely capture the security requirement
that large deviations from the desired properties should be a low
probability event.  
Note that the $m=0$ terms do not contribute, as $\| \, p_{\rm \ss
ideal}^{(m)} - p_{\rm \ss qkd}^{(m)} \, \|_1 \, =\, 0$ and
$I(K_\E\,{:}\,K_\A,K_\B \, | \,M=0\,) = 0$.
%


\section{Quantum Universal Composability Theorem}
\label{sec:qucompos}

Cryptographic protocols often consist of a number of simpler
components.
A single primitive is rarely used alone.  A strong security definition
for the primitive should thus reflect the security of using it within
a larger application.
This allows the security of a complex protocol to be based only on the
security of the components and how they are put together, but not in
terms of the details of the implementation.

A useful approach is to consider the {\em universal composability} 
of cryptographic primitives \cite{Canetti01,QIP03-1,BM02}. 
The first ingredient is to ensure the security of a {\em basic
composition}.
We need a security definition stated for a single execution of the
primitive that still guarantees security of composition with other
systems.
This definition involves a description of some ideal functionality of
the primitive (i.e.~the ideal task the primitive should achieve).
More concretely, we want a security definition such that, if $\s$ is a
secure realization of an ideal subroutine $\s_\I$, and a protocol $\P$
using $\s_\I$, written as $\P{+}\s_\I$, is a secure realization of
$\P_\I$ (the ideal functionality of $\P$), $\P{+}\s$ is also a secure
realization of $\P_\I$.  Throughout the paper, we denote the
associated ideal functionality of a protocol by adding a subscript
$\I$, and we denote a protocol $\P$ calling a subprotocol $\s$ as
$\P{+}\s$ (this last expression stretches the meaning of $\P$ a little
bit to refer to the module of $\P$ calling $\s$). 
The second ingredient is a universal composability theorem stating how a
complex protocol can be built out of secure components.
It is simply a recipe on how to securely perform basic composition
recursively.

\vspace*{1ex}

{\bf The simplifications in analyzing the composable security of QKD}

Our goal is to analyze the unconditional security of QKD by using the
quantum universal composability results in \cite{QIP03-1,BM02}. 
The setting for QKD is simpler than that considered in
\cite{QIP03-1,BM02} in two important aspects.  First, we are only
concerned with unconditional security.
%
%
Second, in QKD, Alice and Bob are known to be honest, and Eve is known
to be adversarial, and there is no unpredicted corruption of any
party.  The formal corruption rules are not used in our derivation of
a composable security definition for QKD.
The following simplified model is sufficient for our derivation of 
a universal composable security definition for QKD.

\vspace*{1ex} 

{\bf The simplified model} 

We first describe the model for quantum protocols and other concepts
involved in the quantum composable security definition.  We base our
discussion on the (acyclic) quantum circuit model (see, for example,
\cite{Yao93a,AKN98}), with an important extension \cite{BM02} (see
also the endnotes \cite{poset}).  Throughout the paper, we only consider 
circuits in the extended model.  \\[0.5ex]
{\it 1. Structure of a protocol~~} A (cryptographic) protocol $\P$ can
be viewed as a quantum circuit in the extended model
\cite{BM02,poset}, consisting of inputs, outputs, a set of registers,
and some partially ordered operations.
A protocol may consist of a number of subprotocols and parties.  
Each subprotocol consists of smaller units called ``unit-roles,'' 
within each the operations are considered ``local.'' 
For example, the operations and registers of each party in each 
subprotocol form a unit-role.  
Communications between unit-roles within a subprotocol represent {\em
internal communications}; those between unit-roles in different 
subprotocols represent input/output of data to the subprotocols.
A channel is modeled by an ordered pair of operations by the sender
and receiver on a shared register.
The channel available to perform each communication determines its
security features.
\\[0.5ex]
{\it 2. The game: security in terms of indistinguishability from the
ideal functionality~~} Let $\P_\I$ denote the ideal functionality of
$\P$.  Intuitively, $\P$ is secure (in a sense defined by $\P_\I$) if
$\P$ and $\P_\I$ behave similarly under any adversarial attack.
%
%
``Similarity'' between $\P$ and $\P_\I$ is modeled by a game between
{\em an environment} $\CE$ and {\em a simulator} $\S$.  These are sets
of registers and operations to be defined, and they are sometimes
personified in our discussion.
In general, $\P$ and $\P_\I$ have very different internal structures
and are very distinguishable, and the simulator $\S$ is added to
$\P_\I$ to make an extended ideal protocol $\P_\I{+}\S$ that is less
distinguishable from $\P$.
$\CE$ consists of the adversaries that act against $\P$ and an
application protocol that calls $\P$ as a subprotocol.
At the beginning of the game, $\P$ or $\P_\I{+}\S$ are picked at
random.
$\CE$ will call and act against the chosen protocol, and will output a
bit $\G$ at the end of the game.
The similarity between $\P$ and $\P_\I{+}\S$ (or the lack of it) is
captured in the statistical difference in the output bit $\G$.
\\[0.5ex]
{\it 3. Valid $\CE$}: The application and adversarial strategy of
$\CE$ are first chosen (the same whether it is interacting with $\P$
or $\P_\I{+}\S$).
$\CE$ has to obey quantum mechanics, but is otherwise unlimited in
computation power.
If $\P$ is chosen in the game, $\CE$ can ({\sc i}) control the
input/output of $\P$, ({\sc ii}) attack insecure internal
communication as allowed by the channel type, ({\sc iii}) direct the
adversarial parties to interact with the honest parties in $\P$.
$\CE{+}\P$ has to be an acyclic circuit in the extended model
\cite{BM02,poset}.
\\[0.5ex]
{\it 4. Valid $\P_\I$ and $\S$}:  
If $\P_\I+\S$ is chosen in the game, $\CE$ ({\sc i}) controls the
input/output of $\P_\I$ as before.
%
%
However, the interaction given by ({\sc ii}) and ({\sc iii}) above will
now occur between $\CE$ and $\S$ instead.  ($\S$ is impersonating or
simulating $\P$.)
The strategy of $\S$ can depend on the strategy of $\CE$.
$\P_\I$ should have the same input/output structure as $\P$, but is
otherwise arbitrary.  (Of course, the security definition is only
useful if $\P_\I$ carries the security features we want to prove for
$\P$.)  
In particular, $\P_\I$ may be defined with internal channels and
adversaries different from those of $\P$.
$\S$ can ({\sc ii}$'$) attack insecure internal communication of $\P_\I$
and ({\sc iii}$'$) direct the adversarial parties to interact with the
honest parties in $\P_\I$.
Thus, $\P_\I$ exchanges information with $\S$, and this can modified
the security features of $\P_\I$.
To $\CE$, $\S$ acts like part of $\P_\I$, ``padding'' it to look
like $\P$, while to $\P_\I$, $\S$ acts like part of $\CE$.
It is amusing to think of $\S$ as making a ``man-in-the-middle'' attack 
between $\CE$ and $\P_\I$.
Finally, $\CE{+}\P_\I{+}\S$ has to be an acyclic circuit in the extended 
circuit model \cite{BM02,poset}. 
See Figure~2 for a summary of the game and the rules.  
%
\\[-2ex]
\begin{figure}[h]
\setlength{\unitlength}{0.4mm}
\centering
\begin{picture}(200,75)
\put(10,0){\framebox(20,30){}}
\put(13,3){\makebox{$\P$}}
\put(10,50){\framebox(60,20){}}
\put(70,59){\vector(1,0){7}}
\put(70,61){\vector(1,0){7}}
\put(76,55){\makebox(10,10){$\G$}}
\qbezier(40,50)(55,75)(70,50)
\put(55,25){\vector(-1,0){30}}
\put(55,25){\vector(0,1){30}}
\put(15,40){\vector(0,1){15}}
\put(15,40){\vector(0,-1){15}}
\put(13,60){\makebox{$\CE$}}
\put(60,36){\makebox(10,5){\sc ii,iii}}
\put(15,37){\makebox(5,5){\sc i}}

\put(115,37){\makebox(5,5){\sc i}}
\put(113,60){\makebox{$\CE$}}
\put(110,0){\framebox(20,30){}}
\put(113,3){\makebox{$\P_\I$}}
\put(110,50){\framebox(60,20){}}
\put(170,59){\vector(1,0){7}}
\put(170,61){\vector(1,0){7}}
\put(176,55){\makebox(10,10){$\G$}}
\qbezier(140,50)(155,75)(170,50)
\put(155,35){\vector(0,1){20}}
\put(155,35){\vector(0,-1){10}}
\put(115,40){\vector(0,1){15}}
\put(115,40){\vector(0,-1){15}}
\put(145,0){\framebox(20,30){}}
\put(157,3){\makebox{$\S$}}
\qbezier(145,12)(158,18)(145,24)
\put(145,18){\vector(-1,0){20}}
\put(145,18){\vector(1,0){5}}
\put(132,19){\makebox(10,5){\sc ii}\hspace*{-0.5ex}$'$}
\put(132,11){\makebox(10,5){\sc iii}\hspace*{-0.3ex}$'$}
\put(160,36){\makebox(10,5){\sc ii,iii}}
\end{picture}
\label{fig:compos}
\caption{The game defining the composable security definition.  The
curved region in $\CE$ represents the adversaries against $\P$, and
the curved region in $\S$ represents the adversaries against $\P_\I$.
We label the types of interactions as described in the text.}
\end{figure}
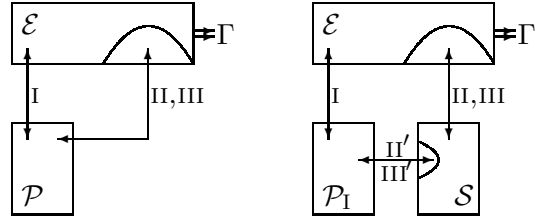

With a slight abuse of language, the symbols $\P$ and $\P_\I{+}\S$ are
also used to denote the respective events of their being chosen at the
beginning of the game.
We can now state the universal composable security definition. \\[1ex]
{\bf Definition 1:} $\P$ is said to $\e$-securely realizes $\P_\I$
(shorthand $\P \; \e$-\sr$\P_\I$) if
%
\bea
	\forall \CE ~~\exists \S  ~~{\rm s.t.}~~ 
	\big | \,\Pr(\G{=}0|\P) - \Pr(\G{=}0|\P_\I{+}\S) \, \big | \leq \e
\;.
\label{eq:usd}
\eea
We call $\e$ in \eq{usd} the {\em distinguishability-advantage} between
$\P$ and $\P_\I$.
This security definition (in the model described) is useful because
security of basic composition follows ``by definition''
\cite{QIP03-1,BM02}.  We have the following simple version of a
universal composability theorem.  \\[1ex]
{\bf Theorem 1:} Suppose a protocol $\P$ calls a subroutine $\s$.  
If $\s$ $\e_\s$-\sr$\s_\I$ and $\P{+}\s_\I$ $e_\P$-\sr$\P_\I$, then 
$\P{+}\s$ $\e$-\sr$\P_\I$ for $\e \leq \e_\P{+}\e_\s$. 


Theorem 1 can be generalized to any arbitrary protocol with a proper
{\em modular structure}.  An example of an improper modular structure
is one with a security deadlock, in which the securities of two
components are interdependent.

Proper modular structures can be characterized as follows. 
Let $\P{+}\s_1{+}\s_2{+}\cdots$ be any arbitrary protocol using a
number of subprotocols.
This can be represented by a $1$-level tree, with $\P$ being the 
parent and $\s_{1,2,\cdots}$ the children.
Each of $\s_{1,2,\cdots}$ may use other subprotocols, and the
corresponding node will be replaced by an appropriate $1$-level
subtree.
This is done recursively, until the highest-level subprotocols (the
leaves) call no other subprotocols.  These are the primitives. 
%
%
It was proved in \cite{BM02} that more general modular structures,
represented by an acyclic directed graph, can be transformed to a
tree.
%
The following composability theorem relates the security of a protocol
$\P$ to the security of all the components in the tree. \\[1ex]
{\bf Theorem 2:} Let $\P$ be a protocol and $T_\P$ the associated
tree.
Let ${\cal M}$ be the protocol corresponding to any node in $T_\P$
with subprotocols ${\cal N}_i$. 
Suppose $\forall {\cal M}$, ${\cal M}{+}{\cal N}_{1\I}{+}{\cal
N}_{2\I}, \cdots,$ $\e_{\cal M}$-\sr${\cal M}_\I$.  
If $\sum_{{\cal M}} \e_{\cal M} \leq \e$, then $\P$ $\e$-\sr$\P^\I$, 
where the sum is over all nodes in $T_\P$. 

Theorem 2 is obtained by recursive use of theorem 1 and the triangle
inequality.  The idea is to replace subprotocols one-by-one by their
ideal functionalities at the highest level, and proceed recursively to
lower levels toward the root.
The distinguishability-advantage between $\P$ and $\P_\I$ is upper
bounded by the sum of all the individual distinguishability-advantages
between pairs of protocols before and after each replacement.
See Figure~4 for an example of $T_\P$ that describes repeated QKD.


Note that the composable security definition for QKD derived in the
simplified setting will remain applicable in the general setting
considered in \cite{QIP03-1,BM02}.  However, when applying Theorem 2
to analyze the security of an application using QKD, one should use a
setting appropriate for that particular application.

In the next section, we analyze QKD in the composability framework.
This is part of our main result, and an example to illustrate the
composability framework.

\section{Universal composable security definition of QKD} 
\label{sec:qkducdef}

We first describe a general QKD scheme in the composability framework.
Then, we tailor an ideal functionality for KD that resembles QKD.
Finally, we express the universal composable security definition of
QKD as a simple distinguishability criteria.  

\subsection{QKD in the game defining security}

Our discussion relies on the existence of authentication schemes that
are universal composable in the quantum setting.  Furthermore, the
authentication scheme should use a key much shorter than the message
to be authentication (so that QKD indeed expands a key).  For example,
the scheme in \cite{Wegman81} satisfies such conditions (composability
is proved in \cite{HLM04}).
Let $\a$ denote any such authentication scheme and let $\a_\I$ denote
ideal authentication.  Let $\k{+}\a$ denote QKD using authentication
scheme $\a$ and let $\k_\I$ denote an ideal KD protocol to be
defined.
By theorem 1, we can focus on the security of $\k{+}\a_\I$, i.e., QKD
using perfectly authenticated classical channels.  
The initial key requirement is embedded in the subroutine $\a_\I$.  
In this case, QKD has no input.  It outputs some bitstrings $k_\A$,
$k_\B$ of certain length $m$ to Alice and Bob, with $m=0$ if and only
if QKD is aborted.  (We can assume that $m$ is publicly announced, and
consider $m$ as an output of QKD.)  
Eve's view (including both quantum and classical data) is given by the state
$\mrho_{\E,k_\A,k_\B}$.

We now turn to the game defining the composable security definition of
QKD.  
Eve is an adversary that is part of the environment $\CE$.  Following
the discussion in \sec{qucompos}, $\CE$ will fix an arbitrary
strategy.  Since there is no input to QKD, the optimal application in
$\CE$ is simply to receive the output keys from $\k{+}\a_\I$ or
$\k_\I$.  $\CE$ will also consist of the action of Eve and other 
circuits that compute $\G$.
%
A schematic diagram is given in Figure 3. 

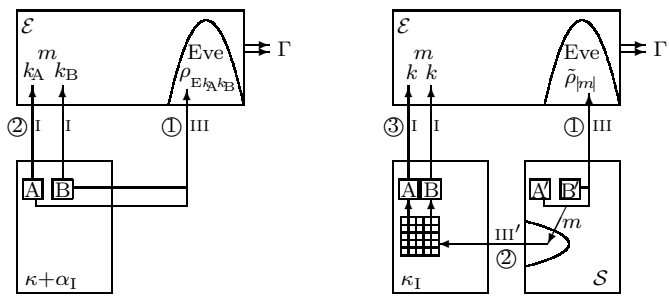
\begin{figure}[h]
\hspace*{-3ex}
\setlength{\unitlength}{0.5mm}
\centering
\begin{picture}(190,70)
{\scriptsize
\put(10,-5){\framebox(25,35){}}
\put(12,-3){\makebox{$\k{+}\a_\I$}}
\put(10,45){\framebox(60,25){}}
\put(70,59){\vector(1,0){7}}
\put(70,61){\vector(1,0){7}}
\put(76,55){\makebox(10,10){$\G$}}
\qbezier(50,45)(60,90)(70,45)
\put(50,54){\makebox(20,10){\scriptsize Eve}}
\put(52,50.5){\makebox(5,5){\scriptsize $\rho$}}
\put(51,46){\makebox(20,9){\tiny $~_{\E  k_{\!\!\A\!} 
					k_{\!\B\!}}$}}
\put(24.5,23){\line(1,0){30.5}}
\put(15,18){\line(0,1){2}}
\put(15,18){\line(1,0){40}}
\put(55,18){\vector(0,1){31}}
\put(14,25){\vector(0,1){25}}
\put(22,25){\vector(0,1){25}}
\put(16,56){\makebox(4,5){\scriptsize $m$}}
\put(10,50){\makebox(18,8){\scriptsize $k_{\!\A}~k_{\ms\B}$}}
\put(11,64){\makebox{$\CE$}}
\put(56,37){\makebox(5,5){\sc iii}}
\put(14.5,37){\makebox(10,5){\sc i~~~i}}
\put(49,37){\makebox(4,5){\scriptsize 1}}
\put(51,39.25){\circle{5}}
\put(8,37){\makebox(4,5){\scriptsize 2}}
\put(10,39.25){\circle{5}}
\put(11.5,20){\framebox(5,5){\scriptsize A}}
\put(19.5,20){\framebox(5,5){\scriptsize B}}

\put(156,37){\makebox(4,5){\scriptsize 1}}
\put(158,39.25){\circle{5}}
\put(108,37){\makebox(4,5){\scriptsize 3}}
\put(110,39.25){\circle{5}}

\put(138,2){\makebox(4,5){\scriptsize 2}}
\put(140,4){\circle{5}}

\put(146.5,20){\framebox(5,5){\scriptsize A\hspace*{-0.5ex}$'$}}
\put(154.5,20){\framebox(5,5){\scriptsize B\hspace*{-0.3ex}$'$}}
\put(159.5,23){\line(1,0){2.5}}
\put(150,18){\line(1,0){12}}
\put(150,18){\line(0,1){2}}

\put(150,47){\makebox(20,9){\scriptsize $\tilde{\rho}_{\ms | \! m \! |}$}}
\put(111.5,20){\framebox(5,5){\scriptsize A}}
\put(117.5,20){\framebox(5,5){\scriptsize B}}
\put(113,37){\makebox(5,5){\sc i}}
\put(119,37){\makebox(5,5){\sc i}}
\put(111,64){\makebox{$\CE$}}
\put(110,-5){\framebox(25,35){}}
\put(112,-3){\makebox{$\k_\I$}}
\put(110,45){\framebox(60,25){}}
\put(170,59){\vector(1,0){7}}
\put(170,61){\vector(1,0){7}}
\put(176,55){\makebox(10,10){$\G$}}
\qbezier(150,45)(160,90)(170,45)
\put(150,54){\makebox(20,10){\scriptsize Eve}}
\put(162,18){\vector(0,1){30}}
\put(114,15){\vector(0,1){5}}
\put(120,15){\vector(0,1){5}}
\put(114,25){\vector(0,1){25}}
\put(120,25){\vector(0,1){25}}
\put(116,56){\makebox(4,5){\scriptsize $m$}}
\put(110,50){\makebox(15,8){\scriptsize $k~k$}}
\put(145,-5){\framebox(25,35){}}
\put(163,-3){\makebox{$\S$}}
\put(156,18){\vector(-1,-2){5}}
\qbezier(145,2)(168,8)(145,14)
\put(151,8){\vector(-1,0){29}}
\put(155,11){\makebox(5,5){\scriptsize $m$}}
\put(137,9){\makebox{{\sc iii}$'$}}
\put(112,5){\framebox(10,10){}}
\multiput(114,5)(2,0){4}{\line(0,1){10}}
\multiput(112,7)(0,2){4}{\line(1,0){10}}
\put(163,37){\makebox(5,5){\sc iii}}
}
\end{picture}
\label{fig:qkdgame}
\vspace*{1ex} 
\caption{The game defining the composable security definition of QKD,
with our choice of ideal KD and simulator.  An ordering of the
interactions is given in circles.  We also label the types of
interactions (see rules 3 and 4 in \sec{qucompos}) explicitly.  Upon
an input $m$, the checkered box generates a perfect key of length $m$
to Alice and Bob.}
\end{figure}

If $\CE$ is interacting with $\k{+}\a_\I$, $\CE$ will: ({\sc i})
receive the output bitstrings $k_\A$, $k_\B$, and $m\,{=}\,|k_\A|
\,{=}\, |k_\B|$, and ({\sc iii}) obtain $\mrho_{\E,k_\A,k_\B}$ which 
depends on Eve's strategy and $k_{\A}$, $k_{\B}$.
Altogether, $\CE$ will be in possession of the state 
\bea
	\mrho_{\rm qkd} = \sum_{k_\A, k_\B} \Pr(k_\A, k_\B) \; 
	|k_\A, k_\B\>\<k_\A, k_\B| \otimes \mrho_{\E,k_\A, k_\B}
\label{eq:rhoreal}
\eea
in which $\mrho_{\E,k_\A,k_\B}$ and $k_\A, k_\B$ can be correlated.
We have omitted an explicit register for $m$, because the information
is redundant given $k_\A, k_\B$.
%
See Figure 3 for a schematic diagram for QKD, and how
it interacts with the environment.

\subsection{Ideal KD and the simulator}
\label{sec:ikd}

We now define the ideal functionality for QKD.  
In general, when formulating an ideal functionality, one need not be
concerned with how the functionality is realized.  
%
%
What is important is to impose the essential security features while
mimicking the analyzed protocol from the point of view of $\CE$.

Our ideal KD functionality $\k_\I$ has to model both the possibility
to generate a perfect key, and the possibility for Eve to cause QKD to
be aborted.
Besides Alice and Bob, $\k_\I$ has a box that accepts a value $m$ from
an adversary ``Devil'' and outputs a perfect $m$-bit key $K$ to Alice
and Bob ($m=0$ means abort).
When $\k_\I$ is run, Devil sends $m$ to the box, which sends 
$K$ to Alice and Bob.
This formulation of $\k_\I$ satisfies the security conditions
\eqs{uniformity}{privacy} perfectly ($\mu_1,\mu_2=0$).
%
See Figure 3 for a schematic diagram.

Consider the following simulator $\S$.
$\S$ runs a ``fake QKD'' with fake Alice$'$ and Bob$'$.  They interact
with Eve (in $\CE$) and run verification procedure as in QKD.  A value
$m$ is announced for the fake QKD, but the fake output keys are unused
and kept secret in $\S$.
The Devil in $\S$ then sends $m$ to the box in $\k_\I$, which
generates a perfect $m$-bit key string $k$ to Alice and Bob in
$\k_\I$, who forward their outputs to $\CE$.
Let 
\bea 
 	\tilde{\mrho}_{m} = 
 	\sum_{k_\A,k_\B:|k_\A|=|k_\B|=m} 
 	\Pr(k_\A, k_\B|M{=}m) \; \mrho_{\E,k_\A, k_\B}  \,.
\label{eq:rhotilde}
\eea
Then, at the end of the game, $\CE$ will be in possession of the state
\bea
	\mrho_{\rm ideal} 
	= \sum_k \Pr(M{=}|k|) \; 2^{-|k|} \; 
	|k,k\>\<k,k| \; {\ot} \; \tilde{\mrho}_{|k|}  \,.
\label{eq:rhoideal}
\eea
See Figure 3 for a schematic diagram for how $\k_\I+\S$ interacts 
with $\CE$.
%
%

\subsection{Universal composable security definition and simple 
privacy condition}
\label{sec:diff}

Recall that at the beginning of the game, one of $\k$ and $\k_\I{+}\S$
is chosen at random to interact with $\CE$.  The
distinguishability-advantage is upper bounded by the trace distance of
the two possible final states of $\CE$ right before $\G$ is computed,
\bea
        \hspace{-2ex} \big | \, \Pr(\G{=}0 \,|\, \k) 
	- \Pr(\G{=}0 \,|\, \k_\I{+}\S) \, \big | 
	& \leq & \smfrac{1}{2} \; 
	\big \| \, \mrho_{\rm qkd} - \mrho_{\rm ideal} \, \big \|_1 
\label{eq:qkdseccond}
\\
	& \leq & \smfrac{1}{2} \;
	\big \| \, \mrho_{\rm qkd} - \mrho_{\rm qi1} \, \big \|_1
	+ \smfrac{1}{2} \;
	\big \| \, \mrho_{\rm qi1} - \mrho_{\rm qi2} \, \big \|_1 
	+ \smfrac{1}{2} \;
	\big \| \, \mrho_{\rm qi2} - \mrho_{\rm ideal} \, \big \|_1 \,, 
\label{eq:qkdseccond2}
\eea
%
%
where $\mrho_{\rm qi1}$ and $\mrho_{\rm qi2}$ are hybrid, intermediate, 
states between $\mrho_{\rm qkd}$ and $\mrho_{\rm ideal}$ defined as
\bea
	\mrho_{\rm qi1} & = &  
	\sum_k \Pr(M{=}|k|) \; 2^{-|k|} |k,k\>\<k,k|
			             \; \ot \; \mrho_{\E,k,k} 
\,, 
\label{eq:rhointer1}
\\
	\mrho_{\rm qi2} 
	& = & \sum_k \Pr(M{=}|k|) \; 2^{-|k|} \; 
	|k,k\>\<k,k| \; {\ot} \; \bar{\mrho}_{|k|} 
\,,~~~{\rm with} 
\label{eq:rhointer2}
\\
	\bar{\mrho}_{m} & = &  
  	\frac{1}{2^m} \sum_{k:|k|=m} 
 	\mrho_{\E,k,k}
\,.
\label{eq:rhobar}
\eea
The sum of the first and the last terms in \eq{qkdseccond2} can be
bounded by $\mu_1$ in the equality-and-uniformity condition
(\eq{uniformity} in \sec{qkd}) as follows.
Using \eqs{rhoreal}{rhointer1}, 
%
{\small \vspace*{2ex} 
\bea
	\big \| \, \mrho_{\rm qkd} - \mrho_{\rm qi1} \, \big \|_1
	= \bigg \| \sum_{k_{\!\A}{\neq}k_\B} 
	    \! \Pr(k_{\!\A}, \!  k_\B) \, 
	    |k_{\!\A}, \!  k_\B\>\<k_{\!\A}, \!  k_\B| 
	    \ox \mrho_{\E,k_\A,k_\B}
	 + \sum_{k} \left[ \Pr(k \ms , \! k){-}\Pr(|k|) \ss 2^{-|k|} \right] 
	    |k\ms , \! k\>\<k\ms , \! k| \ox \mrho_{\E,k,k} \, \bigg \|_1
	\leq  \mu_1 \,.
\non
\eea
}
Using \eqs{rhoideal}{rhointer2}, 
\bea
	\big \| \, \mrho_{\rm qi2} - \mrho_{\rm ideal} \, \big \|_1
	\; \leq \;   
	\sum_{m} 
	 \Pr(M{=}m) \; 
	 \big \| \, \bar{\mrho}_{m} - \tilde{\mrho}_{m} \big \|_1
	\; \leq \; \mu_1
\non
\eea
where we have used  
$\bar{\mrho}_{m} = \sum_{k_\A,k_\B} p^{(m)}_{\rm ideal}(k_\A, k_\B) 
	\, \mrho_{\E,k_\A,k_\B}$, 
$\tilde{\mrho}_{m} = \sum_{k_\A,k_\B} p^{(m)}_{\rm qkd}(k_\A, k_\B) 
	\, \mrho_{\E,k_\A,k_\B}$, 
and the equality-and-uniformity condition \eq{uniformity} for the last
inequality.
The remaining term in the composable security condition
\eq{qkdseccond2} is given by
\bea
	\frac{1}{2} \, 
	\big \| \, \mrho_{\rm qi1} - \mrho_{\rm qi2} \, \big \|_1
	&=& \frac{1}{2} \, \bigg \| \sum_{k} 
	 \Pr(M{=}|k|) \; 2^{-|k|} \; 
	 |k \ms , \! k\>\<k \ms , \! k| \ox 
	 \lbm \! \bar{\mrho}_{|k|} - \mrho_{\E,k,k} \! \rbm \bigg \|_1
\non
\\ 
	&\leq & \frac{1}{2} \, 
	\sum_{k} 
	 \Pr(M{=}|k|) \; 2^{-|k|} \; 
	 \big \| \, \bar{\mrho}_{|k|} - \mrho_{\E,k,k} \big \|_1 \,, 
\label{eq:newseccond}
\eea
which can be interpreted as a new privacy condition.

We have thus compartmentalized the composable security definition for
QKD, \eq{qkdseccond} or \eq{qkdseccond2}, into two parts: the original
equality-and-uniformity condition \eq{uniformity} and a new privacy condition
\eq{newseccond}, which we loosely call a ``composable privacy
condition'' for QKD.
Once \eq{newseccond} is bounded by some $\mu_2^*$, QKD using ideal
authentication $\k{+}\a_\I$ $\e_\k$-securely realizes the ideal KD 
$\k_\I$, if $\mu_1 + \mu_2^* \leq \e_\k$.
Following Theorems 1 and 2, one can use the key ``as if it were
perfect.''
Proving such a bound on \eq{newseccond} is relatively straightforward,
as compared to a direct proof of the security of using a slightly
imperfect key from QKD (without the composability theorem).

In the following section, we prove several bounds for \eq{newseccond}.
First, we show that for any QKD scheme satisfying the usual privacy
condition \eq{privacy}, \eq{newseccond} can be bounded as well, albeit
with a potentially large but manageable degradation.
Second, we prove a tighter bound on \eq{newseccond} assuming a privacy
condition in terms of Eve's Holevo information on the key.
Finally, we propose a new, tight, sufficient condition for bounding
\eq{qkdseccond} (the full composable security condition) (bypassing
\eq{privacy} and automatically incorporating all of equality,
uniformity, and privacy) based on the singlet-fidelity considered in
most existing security proofs for QKD.
As an application, we obtain sharp upper bounds for \eq{qkdseccond} 
for existing QKD schemes.

\section{Universal composability of QKD}
\label{sec:qkduc}

\subsection{Usual privacy condition implies composable privacy condition}

{\bf Bound 1:}~~ We first mention a loose upper bound for 
$\| \mrho_{\rm qi1} - \mrho_{\rm qi2} \|_1$.  It is upper bounded by: 
\bea
	\sum_{m} \Pr(M{=}m) \; \bigg \|  \hspace{-2ex} \sum_{~~~k:|k|=m} 
	 \hspace{-2ex} 2^{-m} \; 
	 |k \ms , \! k\>\<k \ms , \! k| \ox 
	 \lbm \! \bar{\mrho}_{|k|} - \mrho_{\E,k,k} \! \rbm \bigg \|_1
\non
\eea
and according to Lemma 1 of \cite{DiVincenzo03}, each trace distance
is upper-bounded by $(2^m{+}1)^2 \sqrt{2 (\ln 2) \, I(K_\E{:}K|m)}$
(we use the shorthand $m$ for $M=m$ in the mutual information).
Thus 
\bea
	\big \| \mrho_{\rm qi1} - \mrho_{\rm qi2} \, \big \|_1
	&\leq &  \!\! \sum_{m} \Pr(M{=}m) (2^m{+}1)^2 \sqrt{2 (\ln 2) \,
	I(K_\E{:}K|m)}
\non
\\
	&\leq & \!\!  (2^{\max(m)}{+}1)^2 \! \sqrt{2 (\ln 2)}  
	 \, \sum_{m} \! \Pr(\!M{=}m) \sqrt{\!I(K_\E{:}K|m)}
\non
\\[-1ex]
	&\leq &  \!\! (2^{\max(m)}{+}1)^2 \! \sqrt{2 (\ln 2)}  
	\, \lbL \sum_{m} \! \Pr(\!M{=}m) \, I(K_\E{:}K|m) 
	\rbL^{\smfrac{1}{2}}
\non
\\
	&\leq &  \!\! (2^{\max(m)}{+}1)^2 \! \sqrt{2 (\ln 2)} 
         \; \sqrt{\mu_2}
\non
\eea
where the second last line is obtained by the Cauchy-Schwarz
inequality.  Typically, $\max(m)$ is a small fraction of $n$, the
security parameter such as the number of qubits communicated.  Recall
that $\mu_2 \in {\cal V}$, the set of exponentially decaying functions
of $n$.  With a limit on the key rate $m/n$ (based on how fast $\mu_2$
vanishes), $\e_\k \in {\cal V}$ also.  We now derive a slightly better
bound.

{\bf Bound 2:}~~ 
The second bound of \eq{newseccond} requires two lemmas. 
The {\em Shannon distinguishability}~\cite{Fuchs97} of two quantum
states $\r_0$ and $\r_1$, $\SD(\r_0,\r_1)$, is defined as the 
accessible information on $C$ obtained by measuring a specimen of
$\r_C$, where $C$ is a coin toss 
(see \cite{Fuchs97}). \\[2ex]
%
{\bf Lemma 1:} Let $I_{\rm acc}$ be the accessible information of an
ensemble $\{q_x,\r_x\}_{i=1}^{2^m}$ of finite dimensional states
(i.e. $I_{\rm acc}$ is the maximum information obtained on $X$ 
by measuring a single specimen of $\r_X$, where 
$\Pr(X=x)=q_x$).  Let $\r = \sum_x q_x \r_x$.  Then,  
$\forall x$, $q_x \, \SD(\r_x, \r) \leq I_{\rm acc}$. \\[1ex]
{\bf Proof of Lemma 1:} Define random variables $C,X_1,X_2$ and $Y$ as:
\\ 1. $C$ is a coin toss. 
\\ 2. $X_1 = x$ with probability $q_x$, 
\\ 3. If $C = 0$, $X_2 = x$, else $X_2 = x'$ with probability $q_{x'}$.
\\ 4. $Y$ is the outcome of measuring 
      \raisebox{0.3ex}{$\r$}$_{\! X_{\!2}}$.  

These random variables are defined so that for each $x$, $I(Y \: C|X_1
\,{=}\, x)$ is the information gained on whether a randomly drawn
state is $\r_x$ or $\r$ by measuring the state.  Also, $\Pr(X_2
\,{=}\, x) = q_x$ and $\r_{X_2}$ is simply a draw from the initial
ensemble.

Note that $Y$ depends on the measurement.  For the measurement
attaining $\SD(\r_x, \r)$,
\bea
	I(Y \: C|X_1) = \sum_l q_l \, I(Y \: C| \,X_1{=}l) 
 	\geq q_x \,\SD(\r_x,\r) \,,
\eea
whereas for any measurement, 
\bea
	\hspace*{-5ex} I(Y \: C|X_1) & = & I(Y \: X_1 C) - I(Y \: X_1) 
\non
\\	& \leq & I(Y \: X_1 C)
	\leq I(Y \: X_2)
	\leq I_{\rm acc}
\eea
where the three inequalities are respectively due to the Chain rule,
the fact $X_1 C \ra X_2 \ra Y$ is a Markov chain, and the optimality
of $I_{\rm acc}$. \hfill $\square$

%

We also use the following relation between the trace distance and the
Shannon distinguishability, readily obtained from Eq.~(47) and Fig.~1
of \cite{Fuchs97}. \\[2ex] 
{\bf Lemma 2:~~} $\forall \r_0, \r_1$, ~
$ \|\r_0 - \r_1 \|_1 \leq 2 \sqrt{\SD(\r_0,\r_1)}$.  

\vspace{1ex} 

{\bf Proof of bound 2:~~} 
For each key length $m$, define ${\cal F}_m$ to be the ensemble $\{2^{-m},
\mrho_{\E,k,k}\}_{|k|=m}$.
%
Lemmas 1 and 2 imply 
\[
	\big \| \bar{\mrho}_{|k|} - \mrho_{\E,k,k}
	\big \|_1 \leq 2^{\smfrac{m}{2}{+}1} \sqrt{I(K_\E\,{:}\,K|\,m)}
\]
\hfill\\
and we can bound \eq{newseccond} using the above, the 
Cauchy-Schwarz inequality, and the usual privacy condition \eq{privacy}: 
\bea 
	\big \| \mrho_{\rm qi1} - \mrho_{\rm qi2} \, \big \|_1
	&\leq &  \sum_{k} \Pr(M{=}|k|) \; 2^{-|k|} \; 
	 \big \| \, \bar{\mrho}_{|k|} - \mrho_{\E,k,k} \big \|_1
\non
\\ 
	&\leq& \sum_{m} \Pr(M{=}m) \; 2^{\smfrac{m}{2}{+}1} 
	\sqrt{I(K_\E\,{:}\,K|\,m)}
\non
\\ 
	&\leq& 2^{\max(m)/2{+}1} 
	\lbL \sum_{m} \Pr(M{=}m) \; I(K_\E\,{:}\,K|\,m) \rbL^{\smfrac{1}{2}}
\non
\\ 
	&\leq& 2^{\max(m)/2{+}1} \sqrt{\mu_2}
\non
\eea
Once again, the key length is a fraction of $n$, and if appropriate
limits on the key rate are imposed (depending on $\mu_2$), the above
still vanishes exponentially with $n$.

\subsection{Small Holevo information implies composable privacy}

Suppose, instead of the usual privacy condition \eq{privacy} in terms
of the accessible information, we have \\[2ex]
{\it $\bullet$~Privacy:} $\exists \mu_2' \in {\cal V}$ 
s.t.~
\bea 
	\sum_m \Pr(M{=}m) \times \chi({\cal F}_m) ~\leq~ \mu_2'
\label{eq:holevo}
\eea
where $\chi$ is the Holevo information \cite{Holevo73c}, and ${\cal
F}_m = \{2^{-m},\mrho_{\E,k,k}\}_{|k|=m}$ is as defined before.
%
%
\eq{holevo} is more stringent than \eq{privacy} since the Holevo
information is an upper bound for the accessible information.
It was proved in \cite{Schumacher00} that the Holevo information for
an ensemble is the average of the relative entropies $S(\cdot\|\cdot)$
of the states in the ensemble to the average state.
Applying this fact to ${\cal F}_m$, 
\bea 
	\chi({\cal F}_m) = \frac{1}{2^{m}} \sum_{k:|k|=m} 
		S( \mrho_{\E,k,k} \, \| \, \bar{\mrho}_{m} ) \,.
\non 
\eea
Furthermore, the relative entropy is related to the trace distance 
\cite{Ohya93},
\bea
	\big \| \, \mrho_{\E,k,k} - \bar{\mrho}_{m} \, \big \|_1^2 \leq
  	2 \, (\ln 2) \, S(\mrho_{\E,k,k} \, \| \, \bar{\mrho}_{m}) \,. 
\eea
Thus \eq{newseccond} can be bound as 
\bea 
	\big \| \mrho_{\rm qi1} - \mrho_{\rm qi2} \, \big \|_1
	&\leq &  \sum_{k} \Pr(M{=}|k|) \; 2^{-|k|} \; 
	 \big \| \, \bar{\mrho}_{|k|} - \mrho_{\E,k,k} \big \|_1
\non
\\ 
	&\leq & \lbL \sum_{k} \Pr(M{=}|k|) \; 2^{-|k|} \; 
	 \big \| \, \bar{\mrho}_{|k|} - \mrho_{\E,k,k} \big \|_1^2 
	 \rbL^{\smfrac{1}{2}}
\non
\\ 
	& \leq & \lbL 2 \, (\ln 2) \, \sum_{k} \Pr(M{=}|k|) \; 2^{-|k|} \; 
	S(\mrho_{\E,k,k} \, \| \, \bar{\mrho}_{|k|}) 
	 \rbL^{\smfrac{1}{2}}
\non
\\ 
	& \leq & \lbL 2 \, (\ln 2) \, \sum_{m} \Pr(M{=}m) 
	\, \chi({\cal F}_m)
	 \rbL^{\smfrac{1}{2}}
\non
\\ 
	&\leq& \sqrt{2 \, (\ln 2) \; \mu_2'}
\non
\eea
which does not have an overhead exponential in the length of the key 
generated.  

\subsection{A new sufficient condition for composable security}
\label{sec:specificqkd}

We can easily analyze the composable security of any QKD scheme that
has a security proof based on entanglement purification protocol.
All existing QKD schemes have such security proofs.  
The final keys $K_\A$, $K_\B$ are outcomes of Alice and Bob's
measurements on a shared state $\mrho_{\A\B}^m$ for some $m$, and
$\mrho_{\A\B}^m$ is supposed to be $\Phi^{\ot m}$ in the absence of
eavesdropping.  Here, $m$ is again the key length and $\Phi =
\smfrac{1}{2} (|00\>+|11\>)(\<00|+\<11|)$.  The usual privacy
condition \eq{privacy} is obtained by showing the following. \\[2ex]
{\it $\bullet$~High fidelity:} $\exists \mu_2'' \in {\cal V}$ 
s.t.~
\bea
        \sum_{m} \Pr(m)  
        \lbm 1 - F(\mrho_{\A\B}^m,\Phi^{\ot m}) \rbm \leq \mu_2''
\label{eq:dumfid}
\eea
(See \sec{motivation} for the definition of $F$.)  The above 
turns out to provide a sharp bound on \eq{newseccond}, as shown 
below. 

Let $\mrho_{\A\B\E}^m$ be the state held by Alice, Bob, and Eve right
before the final measurements of Alice and Bob.  We only need to
consider $m>0$.
Let $|\psi^m_1\>$ be a purification of $\mrho_{\A\B\E}^m$ on systems
$\AA$, $\BB$, $\EE$ and $\XX$. $|\psi^m_1\>$ is also a purification of
$\mrho_{\A\B}^m$.
By Ulhmann's Theorem \cite{Uhlmann76}, there exists a purification 
$|\psi^m_2\>$ over systems $\AA$, $\BB$, $\EE$ and
$\XX$ such that 
\bea
F(|\psi^m_1\>, |\psi^m_2\>) = F(\mrho_{\A\B}^m,\Phi^{\ot m}) \,. 
\non
\eea
By construction of $|\psi^m_{1}\>$ and $|\psi^m_{2}\>$, measuring
$\AA$ and $\BB$ and tracing $\XX$ results in $\mrho_{\rm qkd}^m$ and
$\mrho_{\rm ideal}^m$ respectively.  But measuring and tracing can
only increase the fidelity of two states.  Thus
\bea
	 F(\mrho_{\rm qkd}^m, \mrho_{\rm ideal}^m) 
	> F(\mrho_{\A\B}^m,\Phi^{\ot m})  \,.
\non
\eea 
Finally, we use the fact 
\[ \big \| \mrho_{\rm qkd}^m - \mrho_{\rm ideal}^m \big \|_1 
  \leq 2 \sqrt{1-F(\mrho_{\rm qkd}^m, \mrho_{\rm ideal}^m)} \]
to obtain 
\bea
	\big \| \mrho_{\rm qkd}^m - \mrho_{\rm ideal}^m \big \|_1^2  
  	\leq 4 \lbm 1 - F(\mrho_{\A\B}^m,\Phi^{\ot m}) \rbm \,. 
\non
\eea
Putting all these together, we can bound \eq{qkdseccond} as 
\bea
	\frac{1}{2} \, 
	\big \| \, \mrho_{\rm qkd} - \mrho_{\rm ideal} \, \big \|_1
	& \leq & \frac{1}{2} \, \sum_{m} 
     \Pr(M{=}m) \; \big \| \mrho_{\rm qkd}^m - \mrho_{\rm ideal}^m \big \|_1 
\non
\\
	&\leq & \frac{1}{2} \, \lbL 
	\sum_{m} \Pr(M{=}m) \;  
	 \big \| \mrho_{\rm qkd}^m - \mrho_{\rm ideal}^m \big \|_1^2
	\rbL^{\smfrac{1}{2}} \leq  \sqrt{\mu_2''} \,.
\end{eqnarray}
\eq{dumfid} is a good new sufficient condition for {\em composable
security}, being part of the standard QKD proof and a tight bound on
\eq{qkdseccond} simultaneously.  It also implies {\em both}
equality-and-uniformity and privacy (unlike a bound on Holevo
information or mutual information which only implies the composable
privacy condition).

\section{Discussions and applications} 
\label{sec:apps}

We have motivated this work with a discussion of the potential gap
between the desired security of using a key generated by QKD and the
security promised by the privacy condition \eq{privacy} used in the
study of ``unconditional security'' of QKD.
Then, we apply the universal composability theorem to obtain a new
security condition that will guarantee the security of using a key
generated from QKD.  
We propose a new privacy condition \eq{newseccond} that is composable,
and useful sufficient conditions such as \eq{holevo} or \eq{dumfid}.
Most interesting of all, we show that a bound on the singlet-fidelity
\eq{dumfid} directly implies the composable security condition
\eq{qkdseccond}.
These are our main contributions (in the context of cryptography).

We also provide a proof that the existing privacy condition
\eq{privacy} does imply \eq{newseccond}, albeit with a degradation
factor in the security exponential in the key size.  This ensures the
security of using a key generated from any QKD scheme that has been
proved secure, provided the key rate is limited accordingly.  Despite
the existence of such connections, we emphasize that future research
should address \eq{qkdseccond}, \eq{newseccond}, \eq{holevo}, or
\eq{dumfid} directly.  We also provide a sharp bound on
\eq{newseccond} based on Holevo's information \eq{holevo} or
singlet-fidelity \eq{dumfid}.  We are glad to find that the existing
security proofs for QKD imply sharp bounds on \eq{qkdseccond}, when
bypassing the usual privacy condition \eq{privacy}.
Outside the context of cryptography, these connections between various
privacy conditions can be useful for the study of correlations in 
quantum systems. 

It is open whether the degradation of the security (that is
exponential in the generated key size) when going from \eq{privacy} to
\eq{newseccond} is necessary.
However, it is a tempting conjecture, as suggested by the pathologies
of the accessible information exhibited recently
\cite{DiVincenzo03,HLSW03}.

As a final application, we analyze the security of repeating QKD $t$
times, without assuming the availability of an authenticated classical
channel.  (Note that $t$ is a fixed parameter that does not grow with
the problem size.) Each run of QKD $\k$ calls a composable
authentication scheme $\a$ as a subroutine, and each run of $\a$
requires a composably secure key, which is provided by the previous
round of $\k$ (as a subroutine to $\a$).  Call the $t$ rounds of QKD
our protocol $\P$.  The associated tree for $\P$, and the ideal
realization $\P_\I$ are given in the far left and right of
Figure~4. 
%
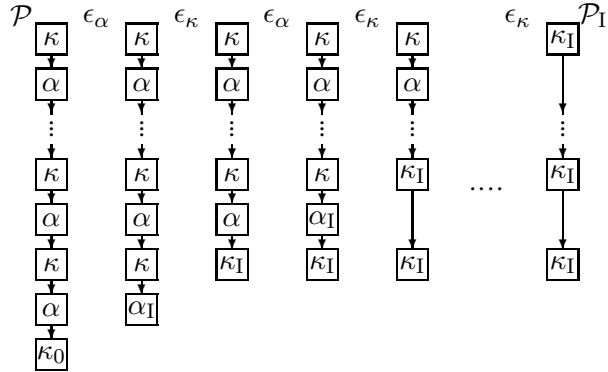
\begin{figure}[http]
\centering
\setlength{\unitlength}{0.4mm}
\begin{picture}(200,120)

\put(0,113){\makebox(10,10){$\P$}}
\put(10,105){\framebox(10,10){$\k$}}
\put(15,105){\vector(0,-1){5}}
\put(10,90){\framebox(10,10){$\a$}}
\put(15,90){\vector(0,-1){5}}
\put(10,77){\makebox(10,10){$\cdot$}}
\put(10,75){\makebox(10,10){$\cdot$}}
\put(10,73){\makebox(10,10){$\cdot$}}
\put(15,75){\vector(0,-1){5}}
\put(10,60){\framebox(10,10){$\k$}}
\put(15,60){\vector(0,-1){5}}
\put(10,45){\framebox(10,10){$\a$}}
\put(15,45){\vector(0,-1){5}}
\put(10,30){\framebox(10,10){$\k$}}
\put(15,30){\vector(0,-1){5}}
\put(10,15){\framebox(10,10){$\a$}}
\put(15,15){\vector(0,-1){5}}
\put(10,0){\framebox(10,10){$\k_0$}}

\put(20,112){\makebox(20,10){$\e_\a$}}

\put(40,105){\framebox(10,10){$\k$}}
\put(45,105){\vector(0,-1){5}}
\put(40,90){\framebox(10,10){$\a$}}
\put(45,90){\vector(0,-1){5}}
\put(40,77){\makebox(10,10){$\cdot$}}
\put(40,75){\makebox(10,10){$\cdot$}}
\put(40,73){\makebox(10,10){$\cdot$}}
\put(45,75){\vector(0,-1){5}}
\put(40,60){\framebox(10,10){$\k$}}
\put(45,60){\vector(0,-1){5}}
\put(40,45){\framebox(10,10){$\a$}}
\put(45,45){\vector(0,-1){5}}
\put(40,30){\framebox(10,10){$\k$}}
\put(45,30){\vector(0,-1){5}}
\put(40,15){\framebox(10,10){$\a_\I$}}

\put(50,112){\makebox(20,10){$\e_\k$}}

\put(70,105){\framebox(10,10){$\k$}}
\put(75,105){\vector(0,-1){5}}
\put(70,90){\framebox(10,10){$\a$}}
\put(75,90){\vector(0,-1){5}}
\put(70,77){\makebox(10,10){$\cdot$}}
\put(70,75){\makebox(10,10){$\cdot$}}
\put(70,73){\makebox(10,10){$\cdot$}}
\put(75,75){\vector(0,-1){5}}
\put(70,60){\framebox(10,10){$\k$}}
\put(75,60){\vector(0,-1){5}}
\put(70,45){\framebox(10,10){$\a$}}
\put(75,45){\vector(0,-1){5}}
\put(70,30){\framebox(10,10){$\k_\I$}}

\put(80,112){\makebox(20,10){$\e_\a$}}

\put(100,105){\framebox(10,10){$\k$}}
\put(105,105){\vector(0,-1){5}}
\put(100,90){\framebox(10,10){$\a$}}
\put(105,90){\vector(0,-1){5}}
\put(100,77){\makebox(10,10){$\cdot$}}
\put(100,75){\makebox(10,10){$\cdot$}}
\put(100,73){\makebox(10,10){$\cdot$}}
\put(105,75){\vector(0,-1){5}}
\put(100,60){\framebox(10,10){$\k$}}
\put(105,60){\vector(0,-1){5}}
\put(100,45){\framebox(10,10){$\a_\I$}}
\put(105,45){\vector(0,-1){5}}
\put(100,30){\framebox(10,10){$\k_\I$}}

\put(110,112){\makebox(20,10){$\e_\k$}}

\put(130,105){\framebox(10,10){$\k$}}
\put(135,105){\vector(0,-1){5}}
\put(130,90){\framebox(10,10){$\a$}}
\put(135,90){\vector(0,-1){5}}
\put(130,77){\makebox(10,10){$\cdot$}}
\put(130,75){\makebox(10,10){$\cdot$}}
\put(130,73){\makebox(10,10){$\cdot$}}
\put(135,75){\vector(0,-1){5}}
\put(130,60){\framebox(10,10){$\k_\I$}}
\put(135,60){\vector(0,-1){20}}
%
\put(130,30){\framebox(10,10){$\k_\I$}}

\put(180,105){\framebox(10,10){$\k_\I$}}
\put(185,105){\vector(0,-1){20}}
%
\put(180,77){\makebox(10,10){$\cdot$}}
\put(180,75){\makebox(10,10){$\cdot$}}
\put(180,73){\makebox(10,10){$\cdot$}}
\put(185,75){\vector(0,-1){5}}
\put(180,60){\framebox(10,10){$\k_\I$}}
\put(185,60){\vector(0,-1){20}}
%
\put(180,30){\framebox(10,10){$\k_\I$}}

\put(160,112){\makebox(20,10){$\e_\k$}}

\put(150,55){\makebox(10,10){$\cdot$}}
\put(153,55){\makebox(10,10){$\cdot$}}
\put(156,55){\makebox(10,10){$\cdot$}}
\put(159,55){\makebox(10,10){$\cdot$}}

\put(190,113){\makebox(10,10){$\P_\I$}}

\end{picture}
\label{fig:tqkd}
\caption{Associated tree for $t$ rounds of $\k$ in the left.  $\k_0$
represents some initially shared key.  The arrows point from parents
to children.  Each tree to the right is obtained by replacing one node
by its ideal functionality.  The distinguishability-advantage of each
pair of consecutive schemes is marked between their trees near the
roots.  Authentication is omitted in the ideal functionality $\P_\I$.}
\end{figure}
 
If $\k{+}\a_\I$ $\e_\k$-\sr$\k_\I$ (as in \eq{qkdseccond}) and if
$\a{+}\k_\I$ $\e_\a$-\sr$\a_\I$, $\P\;$ $t(\e_\k{+}\e_\a)$-\sr$\P_\I$.  In
other words, each additional around of QKD degrades the security
parameter by an additive constant $(\e_\k+\e_\a)$.  The same result
can be obtained by using Theorem 2, or conversely, this simple
exercise illustrates the idea behind Theorem 2.

\section{Acknowledgements}

We thank Charles Bennett, Daniel Gottesman, Aram Harrow, and John
Smolin for interesting discussions on the security concerns of using a
key obtained from QKD.  We also thank Dominique Unruh and J\"orn
M\"uller-Quade for interesting discussions on their alternative
framework of composability.
%

Part of this work was completed while MH and JO were visiting the MSRI
program on quantum information, Berkeley, 2002.  MH is supported by EU
grants RESQ (IST-2001-37559), QUPRODIS (IST-2001-38877).
DL acknowledges the support from the Richard Tolman Foundation and the
Croucher Foundation.  DL and DM acknowledge support from the US NSF
under grant no. EIA-0086038.
JO is supported by EU grant PROSECCO (IST-2001-39227) and a grant from
the Cambridge-MIT Institute.

%


\appendix

\section{Notations}

We gather most of the notations used in the paper, roughly in the order 
of first appearance: 
\begin{itemize}
\item 
KD: key distribution
\item
QKD: quantum key distribution
\item 
Alice and Bob: two honest parties trying to establish a common key
\item
Eve: an active adversary 
\item
$\A$, $\B$, $\E$: subscripts labelling objects related to Alice, Bob,
and Eve respectively \\ $\AA$, $\BB$, $\EE$: labels of their
respective quantum systems
\item
Capitalized letters denote random variables and
the corresponding uncapitalized letters denote particular outcomes
\item
$K_\A$, $k_\A$, $K_\B$, $k_\B$: output keys for Alice and Bob
\item
$K$, $k$: $k:=k_\A$ when $k_\A = k_\B$  
\item
$M$, $m$: publicly announced key length at the end of QKD. $M=0$ iff 
QKD is aborted.
\item
$K_\E$, $k_\E$: classical data possibly extracted by Eve at the 
end of QKD by measuring her quantum state
\item
$\Pr(\cdot)$: probability of the event ``$\cdot$''
\item 
$\log$: logarithm in base $2$
\item
$H(X)$, $I(X:Y)$, $I(X:Y|Z)$, and $I(X:Y|Z{=}z)$ for random variables 
$X$, $Y$, $Z$: \\
$H(X) := - \sum_x \Pr(x) \log \Pr(x)$ is the entropy of $X$ \\
$I(X:Y) := H(X) + H(Y) - H(XY)$ is the mutual information between $X$ 
and $Y$ \\
$I(X:Y|Z{=}z)$ is the mutual information between $X$ 
and $Y$ conditioned on $Z=z$ \\
$I(X:Y|Z):= \sum_z \Pr(z) I(X:Y|Z{=}z)$ is the conditional
mutual information
\item
$\mrho$: generic symbol for a density matrix
\item 
$|\cdot\>$, $|\cdot\>\<\cdot|$: $|\cdot\>$ denotes a vector in a
Hilbert space, with label ``$\cdot$''.  $|\cdot\>\<\cdot|$ denotes the
``outer-product'' of $|\cdot\>$ and $\<\cdot|$ or the projector onto
the subspace spanned by $|\cdot\>$.
\item 
$\Tr(\cdot)$: the trace
\item
$\Tr_{\H_1}(\cdot)$: the partial trace over the system $\H_1$.  Let
$\mrho_{12}$ be the density matrix for a joint state on $\H_1$ and
$\H_2$.  $\Tr_{\H_1}(\mrho_{12})$ is the state after $\H_1$ is
discarded.
\item 
$\| \cdot \|_1$: the trace distance, which can be taken as the sum of
the singular values
\item 
$F$: the fidelity.  For two states $\mrho_1, \mrho_2$ in $H$, 
$F(\mrho_1, \mrho_2) =
\max_{|\psi_1\>,|\psi_2\>} |\<\psi_1|\psi_2\>|^2$ where 
$|\psi_{1,2}\> \in \H \ot \H'$ are
``purifications'' of $\mrho_{1,2}$ (i.e., $\Tr_{\H'}
|\psi_{1,2}\>\<\psi_{1,2}| = \mrho_{1,2}$), and $\<\cdot|\cdot\>$ is
the inner product.  
\item 
$\mrho_{\E,k_\A,k_\B}$: Eve's view (both quantum and classical data) 
when the key outputs to Alice and Bob are $k_\A,k_\B$.  
\item 
$n$: security parameter such as the number of qubits communicated in QKD
\item
$p_{\rm \ss qkd}^{(m)}$: the distribution of $K_\A,K_\B$
generated in QKD conditioned on $|K_\A|=|K_\B|=m$, \\
i.e.,
$p_{\rm \ss qkd\,}^{(m)}(k_\A,k_\B) = \Pr(K_\A=k_\A,K_\B=k_\B|M=m)$.
\item 
$p_{\rm ideal}^{(m)}$: the distribution over two $m$-bit
strings defined as 
$p_{\rm \ss ideal\,}^{(m)}(l,l') = 0$ if $l \neq l'$, 
$p_{\rm \ss ideal\,}^{(m)}(l,l) = 2^{-m}$.  
\item 
${\cal V}$: 
the set of exponentially decaying functions of $n$
\item 
$\s$, $\P$, $\s_\I$, $\P_\I$: $\s$ and $\P$ are generic labels for
protocols, with $\s$ possibly used as a subroutine.  The symbol of a
protocol with a subscript $\I$ denotes the ideal functionality of the 
protocol.  
$\P{+}\s$: a protocol $\P$ calling a subroutine $\s$.  
\item 
$\CE$, $\S$: the environment and the simulator.  These are sets of
registers and operations and they are sometimes personified in our
discussion.
\item 
$\G$: output bit of $\CE$ 
\item 
$\e$-\sr: $\P \; \e$-\sr$\P_\I$ is a shorthand for $\P$ $\e$-securely
realizes $\P_\I$ (see mathematical definition in \eq{usd}).  $\e$ is
called the {\em distinguishability-advantage} between $\P$ and
$\P_\I$.
\item 
$T_\P$: the associated tree for a protocol $\P$
\item 
$\a$, $\a_\I$: universal composable authentication with negligible 
key requirement and its ideal functionality
\item 
$\k{+}\a$, $\k{+}\a_\I$, $\k_\I$: 
QKD using authentication $\a$, 
QKD using ideal authentication $\a_\I$, and 
ideal KD defined in \sec{ikd}
\item
Devil: an adversary that determines the key length $m$ generated by
$\k_\I$
\item 
$\mrho_{\rm qkd}$: state possessed by $\CE$ after interacting with 
$\k{+}\a_\I$, see \eq{rhoreal}
\item 
$\mrho_{\rm ideal}$: state possessed by $\CE$ after interacting with 
$\k_\I$, see \eq{rhoideal}
\item 
$\mrho_{\rm qi1}$, $\mrho_{\rm qi2}$: hybrid, intermediate, 
states between $\mrho_{\rm qkd}$ and $\mrho_{\rm ideal}$, see 
\eqs{rhointer1}{rhointer2}
\item 
$\tilde{\mrho}_{m}$: Eve's state when $M=m$, averaged over $K_\A$,
$K_\B$.  See \eq{rhotilde}
\item 
$\bar{\mrho}_{m}$: uniform average of $\mrho_{\E,k,k}$ for $|k|=m$. 
See \eq{rhobar}
\item 
Ensemble: a distribution $\{q_x\}_x$ of quantum states $\varrho_x$
denoted by $\{q_x, \varrho_x\}_x$
\item 
$I_{\rm acc}$: accessible information of an ensemble $\{q_x,
\varrho_x\}_x$, i.e., the maximum mutual information between $X$ and  
outcome $Y$ obtained from measuring a specimen $\varrho_x$
\item
$\SD(\r_0,\r_1)$: Shannon distinguishability of $\r_0$ and $\r_1$,
defined as $I_{\rm acc}$ of the uniform distribution of $\{\r_0, \r_1\}$.
\item 
${\cal F}_m$: the ensemble $\{2^{-m},\mrho_{\E,k,k}\}_{|k|=m}$
\item 
$\chi(\{q_x, \varrho_x\})$: Holevo information of an ensemble, given
by $S(\sum_x q_x \varrho_x) - \sum_x q_x S(\varrho_x)$ where
$S(\cdot)=\Tr(\cdot \log(\cdot))$ is the von Neumann entropy
\item 
$\mrho_{\A\B}^m$: state on which measurements by Alice and Bob 
output $K_\A$, $K_\B$ in QKD-security-proofs based on entanglement 
purification
\item 
$\Phi$: a perfect EPR pair $\smfrac{1}{2} (|00\>+|11\>)(\<00|+\<11|)$ 
\item Singlet fidelity: $F(\mrho_{\A\B}^m,\Phi^{\ot m})$.  Note that 
``singlet'' usually refers to a state that is only unitarily equivalent 
to $\Phi$, but we borrow the term in this paper.

\end{itemize}


\end{document}